\documentstyle[preprint,prl,aps]{revtex}

\begin{document}
\draft

\title{GAPS IN THE HEISENBERG-ISING MODEL}

\author{Rudolf A.\ R\"{o}mer}

\address{Condensed Matter Theory Unit,
Jawaharlal Nehru Centre for Advanced Scientific Research,
Indian Institute of Science Campus,
Bangalore 560012,
India}

%and

\author{Hans-Peter Eckle\cite{presadd}}

\address{Department of Physics,
         Princeton University,
         Princeton, New Jesey 08544, U.S.A.}

%and

\author{Bill Sutherland}

\address{Physics Department, University of Utah, Salt Lake City, Utah
84112, U.S.A.}

\date{Version: March 2, 1995; printed \today}
\maketitle

\begin{abstract}
We report on the closing of gaps in the ground state of the
critical Heisenberg-Ising chain at momentum $\pi$.
For half-filling, the gap closes at special values of the
anisotropy $\Delta= \cos(\pi/Q)$, $Q$ integer.
We explain this behavior with the help of the Bethe Ansatz and show
that the gap scales as a power of the system size
with variable exponent depending on $\Delta$.
We use a finite-size analysis to calculate this exponent in the critical
region, supplemented by perturbation theory at $\Delta\sim 0$.
For rational $1/r$ fillings, the gap is shown to be closed for
{\em all} values of $\Delta$ and the corresponding perturbation
expansion in $\Delta$ shows a remarkable cancellation of various
diagrams.
%Adding a non-integrable next-nearest neighbor interaction, we show
%that these gaps open up again.
\end{abstract}


\narrowtext
%\tighten

%%%%%%%%%%%%%%%%%%%%%%%%%%%%%%%%%%%%%%%%%%%%%%%%%%%%%%%%%%%%%%%%%%%%%%%%
%
% Introduction
%
%%%%%%%%%%%%%%%%%%%%%%%%%%%%%%%%%%%%%%%%%%%%%%%%%%%%%%%%%%%%%%%%%%%%%%%%

\section{Introduction}

In recent years, there has been much interest in the finite-size spectra
of one-dimensional (1D) quantum many-body Hamiltonians. This effort was
largely
motivated by the fascinating ideas of conformal invariance, which allow
for a computation of scaling indices and classification of universality
classes directly from the $1/L$ behavior of a given critical model
\cite{Cardyreview}.

For 1D systems, solvable by the Bethe-Ansatz (BA) method, the finite
size spectra are accessible analytically by various methods and these
models have thus been used as testing grounds for the general ideas
of conformal invariance.
In particular, the Heisenberg-Ising (H-I) model of an anisotropic
spin-$1/2$ chain has been studied in great detail\cite{yy66}.
It has been shown that its low
energy spectrum in the large $N$ limit can be classified as that of a
$c=1$ conformal field theory with marginally irrelevant operators giving
rise to logarithmic corrections\cite{we}.

%However, most studies of the H-I model have been restricted to the
%antiferromagnetic critical region of the model\cite{we,ss90}.
%In the ferromagnetic interaction
%range, solutions of the BA equations for the low lying states correspond
%to complex roots arranged in strings and their behavior becomes
%increasingly complicated as the anisotropy approaches the
%ferromagnetic isotropic point \cite{yf}.

In the present work, we shall reexamine the behavior of the spectrum
and the corresponding states of the H-I chain threaded by a magnetic flux
$\Phi$.
As shown in \cite{yf}, the adiabatic ground state for half-filling
is periodic in $\Phi$ with a period equal to $2\cdot2\pi$
--- $2\pi$ is the flux quantum in appropriate units --- for values
of the interaction strength $\Delta= \cos(\pi/Q)$ with $Q$ not an
integer.
This is due to the presence of a finite gap at $\Phi=2\pi$.
We show that this gap scales as a power of the system size,
with variable exponent depending on $\Delta$.
For $\Delta\sim 0$, we calculate the exponent explicitly by perturbation
theory.
For $Q$ equal to an integer, however, the gap closes exactly and the
periodicity of the ground state becomes macroscopic at half-filling,
i.\ e.\ proportional to the system size.
We show how this behavior can be understood in terms of the BA solution
as a successive crossing of string solutions.

Furthermore, for $1/r$ filling with $r$ an integer, the periodicity is
again not the expected $r$ flux quanta and we observe exact degeneracies
between ground state and first excited state at various values of $\Phi$.
This is true for {\em all} values of the interaction strength and not only
at isolated points.
We show that the numerical results are confirmed by second order
perturbation theory, due to a remarkable cancellation of various diagrams.

Lastly, we demonstrate that all degeneracies can be removed by simply
adding a non integrable next-nearest neighbor term to the Hamiltonian.

%%%%%%%%%%%%%%%%%%%%%%%%%%%%%%%%%%%%%%%%%%%%%%%%%%%%%%%%%%%%%%%%%%%%%%%%
%
% Closing of the gap for half-filling at special values of the
% interaction strength \mu_q
%
%%%%%%%%%%%%%%%%%%%%%%%%%%%%%%%%%%%%%%%%%%%%%%%%%%%%%%%%%%%%%%%%%%%%%%%%

\section{Closing of energy gaps at half-filling}

As is well known, the H-I model has an interpretation as a lattice
gas of either fermions or bosons, where spin-down represents a particle
and spin-up represents an empty site. The magnetic flux $\Phi$ modifies
only the hopping term, s.t.\ the Hamiltonian is
\begin{equation}
H= - \frac{1}{2} \sum_{j=1}^{N}
	\{ e^{i \Phi/N }c_j^{\dagger} c_{j+1} + \mbox{h.c.} \}
	+\cos\mu \sum_{j=1}^{N} n_j n_{j+1} + e_0,
\label{eqn-ham}
\end{equation}
where $e_0 = \Delta ( 2 M - N/2 )$, $N$ is the number of sites,
$M$ the number of particles and we reparametrized
the anisotropy as $\Delta= -\cos(\mu)$. We note that the total momentum
induced into the system by the flux is given as $P= M\Phi/N$.
We further remark, that
the repulsive region $0\leq\mu<\pi/2$ and the attractive
region $\pi/2<\mu\leq\pi$ of the H-I chain are related by an inversion
of the spectrum: The behavior of the low-lying energy states in the attractive
region will be mirrored to the behavior of the high energy states
in the repulsive region and vice versa.

Let $E_0(\Phi)$ denote the ground state energy of the system with flux
$\Phi$. As is easy to see,
$E_0(\Phi)$ has a periodicity of $2\pi$ in $\Phi$, i.e.\
$E_0(\Phi + 2\pi)= E_0(-\Phi) = E_0(\Phi)$.
However, the periodicity of a given state can be an integer multiple
of the period of the energy.
Following \cite{ss90}, we define a topological winding
number $n$ to be the number of times the flux $\Phi$ increases by $2\pi$
before the state returns to its initial value.

We write the shift of the ground state energy as a function of flux as
$\Delta E_0(\Phi)\equiv E_0(\Phi)-E_0(0) \equiv D \Phi^2/2 N + O(\Phi^4)$,
where $D$ has been called the stiffness constant\cite{ss90}.
For $0\leq\mu\leq\pi/2$ and a half-filled band ($M=N/2$), the stiffness has
been calculated exactly as $D=v_s/2(\pi-\mu)$ for flux values
$|\Phi| \leq 2(\pi-\mu)$;
the spin-wave velocity is given by $v_s= \pi \sin\mu / \mu$\cite{yy66}.

At half-filling, we now adiabatically boost the zero momentum ground
state until we have $\Phi=2\pi$ and $P=\pi$.
For $0\leq\mu<\pi/2$, we then observe a finite gap $\Delta E$ between
the boosted ground state and the first excited state in this
momentum $\pi$ sector and so the winding number of the zero momentum
ground state is $n=2$.
For the non-interacting case of $\mu=\pi/2$, the gap closes and
$n=N$ which correctly implies free acceleration in the thermodynamic limit.

For $\mu>\pi/2$, the situation is more complicated:
As has already been noted in \cite{yf}, there are special values
of the interaction strength $\mu_Q= (Q-1)\pi/Q$ at which the winding
number becomes again macroscopic.
In Fig.\ (\ref{fig-halffill}) we show as an example the spectrum at
$\mu_3= 2\pi/3$ and $N=12$, $M=6$.
Note that not only the ground state gap closes at $\Phi=2\pi$,
but also various higher lying states become degenerate.
Moreover, the closing of the gap at the special values
$\mu_Q$ for $Q=2,3,\ldots,N$ occurs for {\em all} lattices sizes $N$.

In order to further understand this interesting behavior of the
energy spectrum, we studied the Bethe Ansatz equations of the
H-I chain with flux. The equations are given as \cite{ss90}
\begin{equation}
N \theta(\alpha,\mu) =
 2 \pi I + \Phi + \sum_{\alpha'}^{M} \theta(\alpha-\alpha', \mu),
\label{eqn-ba}
\end{equation}
where we have used Yang's\cite{yy66} change of variables for the
pseudomomenta such
that $p = \theta(\alpha,\mu)\equiv 2 \arctan[ \cot(\mu/2)
\tanh(\alpha/2) ]$.
The quantum numbers $\{I\}$ are integers or half-odd integers and
specify the states.
The ground state is given by $\{I\}= \{-(M-1)/2, \ldots, (M-1)/2\}$.

Let us briefly recall the behavior of the roots $\{\alpha\}$ for the ground
state as we adiabatically
turn on the flux:
The largest root $\alpha_M$ goes to $\infty$ exactly
at $\Phi=2(\pi-\mu)$ and the remaining $M-1$ roots are distributed
symmetrically around $0$. The energy of the system is now equal to
$M-1$ particles on a chain of length $N$  with $\Phi=0$ ---
the ground state in the $S^z = 1$ sector. For $0\leq\mu\leq\pi/2$,
further increase of $\Phi$ renders $\alpha_M$ complex and it moves
backwards along the line $i\pi$ with decreasing real part until at
$\Phi=2 \pi$ it has the value $\alpha_M= i \pi\equiv\lambda_1$.
Again, the remaining
$M-1$ real roots have been distributed symmetrically around $0$.
However, the energy now is not related to the $S^z=1$ ground state in
a simple way.

For $\pi/2\leq\mu\leq 2\pi/3$ and at $\Phi=2\pi$,
we have two complex roots with zero real part, i.e.\
$\lambda_{1,2} = i (\pi \pm (\pi-\mu) )$, i.e.\ a two-string.
And in general we find that at $\Phi=2\pi$ in the interaction interval
$(Q-1)\pi/Q\leq\mu\leq(Q)\pi/(Q+1)$, we have a $Q$-string sitting
on the imaginary axis symmetrically around $i \pi$ with individual
roots given by
$\lambda_a= i \pi + i \beta_a \equiv i\pi + i(\pi-\mu) (Q+1-2a) + i\delta_L$.
$\delta_L$ represents an exponentially small finite-size correction,
which is exactly zero for a $1$-string, a $2$-string and the two complex
roots closest to $i\pi$ in a $Q$-string with $Q$ even.
The remaining $M-Q$ real roots are distributed symmetrically around $0$.

Note that at values of the interaction strength equal
to $\mu_Q$, both a $(Q-1)$-string and $Q$-string coexist in the degenerate
ground state. The energies associated with these strings are
$E_Q(\mu_Q) = 0$ and $E_{Q-1}(\mu_Q)= 2 (\cos \mu_Q + 1)$ \cite{s94}.
This suggests, that for $(Q-1)\pi/Q\leq\mu\leq(Q)\pi/(Q+1)$, when
the $Q$-string corresponds to the ground state, the first exited state
corresponds to the $(Q+1)$-string.

We have therefore rewritten the BA equations (\ref{eqn-ba})
in order to incorporate the $Q$-strings sitting on the imaginary axis.
The equations now are
\begin{mathletters}
\label{eqn-baq}
\begin{eqnarray}
N \theta_{00}( \alpha, \mu)
 &= &
   2\pi I
   + \sum_{\alpha'}^{M-Q} \theta_{00}(\alpha - \alpha', 2\mu )
   + \sum_{\beta'}^{Q}   \theta_{01}(\alpha, \beta', \mu ), \\
N \theta_{11}( \beta, \pi-\mu )
 &= &
  0
  + \sum_{\alpha'}^{M-Q} \theta_{10}(\alpha', \beta, \mu )
  + \sum_{\beta'}^{Q}   \theta_{11}(\beta - \beta', 2(\pi-\mu) ),
\end{eqnarray}
\end{mathletters}
with $\theta_{00}(\alpha,\mu)\equiv\theta(\alpha,\mu)$ the phase shift
of real roots $\alpha$,
$\theta_{11}(\beta,\mu) \equiv
 \mbox{Re}\{ 2 \mbox{arctanh}[ \tan(\beta/2) / \tan(\mu/2) ] \}$
the phase shift of complex roots and
$\theta_{01}(\alpha,\beta,\mu) \equiv
 \mbox{Re}\{ 2 \arctan[ \coth[ (\alpha-i\beta)/2) / \tan(\mu) ] \}$,
$\theta_{10}(\alpha,\beta,\mu) \equiv
 -\frac{1}{2} \log[ (\cos(\beta+2\mu) + \cosh(\alpha) ) /
                    (\cos(\beta-2\mu) + \cosh(\alpha) ) ]$
the mixed phase shifts of a real with a complex root.
The quantum numbers for the $\{\beta\}$'s are zero and the flux $\Phi=2\pi$
has been incorporated into the $\{I\}$ quantum numbers.
We caution the reader that care has to be taken in regard to the branch
cuts of these phase shifts. The existence of roots on different branches
is equal to a net shift of the quantum numbers and thus a different state.

In Fig.\ (\ref{fig-gap}), we show the results of iterating the
BA equations (\ref{eqn-baq}) in the momentum $\pi$ sector for the ground
state and the first excited state on a chain with $M=4$ particles on
$N=8$ sites.
For $0\leq\mu\leq\pi/2$, the ground state
corresponds to a $1$-string state. As anticipated above, the first
excited state is given by a $2$-string state.
At $\mu_2$, this $2$-string state takes over as ground state and the
$1$-string state has vanished and in fact now corresponds to a higher
lying state. An initially degenerate $3$-string state
appears and becomes the first excited state as $\mu$ is further increased.
At $\mu_3$, the $2$-string now vanishes, the $3$-string state
takes over as ground state and a new $4$-string state emerges.
As there is no $5$-string on a chain with $M=4$ particles, this
exchange of states ends here.

We have plotted the energy difference $\Delta E \equiv E_1 - E_0$ at
$\Phi=2\pi$ as a function of $\mu$ in Fig.\ (\ref{fig-gap}).
The closing of the gap at the special points $\mu_Q$ can be clearly seen.
Note that $\Delta E/E_0$ is already less than $2\%$ at $\mu= 7\pi/12$ for
this small system. For $N=12$ and $M=6$, $\Delta E \sim 10^{-3}$ for the
largest gap at $\mu\sim 7\pi/12$ and decreases approximately
exponentially as $\mu\rightarrow\pi$.

As $N\rightarrow\infty$, the gaps for $\mu\neq\mu_Q$ should close
as $\Delta E \sim N^{-\gamma}$.
%and finite-size arguments of $c=1$ conformal theory predict the scaling
%behavior to be given by $\gamma= \mbox{integer}$.
In Fig.\ (\ref{fig-gamma}), we have plotted $\gamma$ versus $\mu$ as
extrapolated from calculations of up to $N=14$ sites.
Note that we can observe only $Q$-strings up to $Q=7$ for these
small sizes. Thus we can define $\gamma$ only for $\mu<\mu_7= 6\pi/7$.
Moreover, for $\mu\geq\mu_Q$, useful data can come only from systems
with $M=Q+1$ particles.
The result shows that $\gamma$ varies continuously for all
$0\leq\mu\leq\pi$.
Alcaraz et al.\ \cite{we} have argued that both $1$-string and $2$-string
for $0\leq\mu\leq\pi/2$ have the same scaling dimension. Therefore, the
energy gap $\Delta E$ measures finite size behavior beyond simple conformal
$1/N$ formulas and we interpret the continuous variation of $\gamma$
as indicating the presence of logarithmic corrections.
A direct calculation of these corrections in the H-I chain has so far been
done only at $\mu=0$ and we are presently trying to extent these methods.

However, at $\mu\sim\pi/2$ ($\Delta\sim 0$), we can calculate $\gamma$
directly by perturbation theory in $\Delta$.
As a starting point, we choose the plane wave basis of free particles at
$\Phi=0$ and write
\begin{equation}
c_j= \frac{1}{\sqrt N} \sum_{m=1}^{N} e^{i \frac{2\pi}{N} j m} d_m.
\end{equation}
We can then write the hopping term as
\begin{equation}
T = - 2 \sum_{m=1}^{N} \cos( \frac{2\pi}{N} m + \frac{\Phi}{N} )
 d_m^{\dagger} d_m.
\end{equation}
Similarly, the interaction term is given as
\begin{equation}
V = -\frac{2\Delta}{N}
    \sum_{m_1,m_2,m_3,m_4} \cos\left(\frac{2\pi(m_3-m_4)}{N}\right)
    \Delta_{m_1+m_3, m_2+m_4}
    d_{m_1}^{\dagger} d_{m_2}^{\dagger} d_{m_3} d_{m_4},
\end{equation}
where the periodic Kronecker symbol is defined as
$\Delta_{n, m}= 1$ if $n=m\ \mbox{mod}\ N$ and $0$ otherwise.

We now wish to identify the ground state $\Psi_0$ and the first
excited state $\Psi_1$ which are degenerate at $\Phi=2\pi$, i.e.,
\begin{equation}
\Psi_0^{\dagger} T \Psi_0 |_{\Phi=2\pi} =
 \Psi_1^{\dagger} T \Psi_1 |_{\Phi=2\pi}.
\label{eqn-e0e1}
\end{equation}
Let $M$ be odd and consider $\Phi=0$.
We then construct the ground state simply by filling all
available momenta symmetrically, i.e.,
$\Psi_0= \prod_{m=-(M-1)/2}^{(M-1)/2} d_m^{\dagger} |0\rangle$.
Turning on the flux, we find
$\Psi_0^{\dagger} T \Psi_0 |_{\Phi=2\pi} =
 \Psi_0^{\dagger} T \Psi_0 |_{\Phi=0} + 4 \sin (\pi/N )$.
Keeping in mind Eqn.\ (\ref{eqn-e0e1}), we see that the first excited
state is given by
$\Psi_1|_{\Phi=0}= d_A^{\dagger} d_B^{\dagger} d_C d_D \Psi_0|_{\Phi=0}$
with $A= -(M+3)/2$, $B= -(M+1)/2$, $C=(M-1)/2$ and $D=(M-3)/2$.
A simple calculation shows that Eqn.\ (\ref{eqn-e0e1}) holds indeed.

For $\Delta\neq 0$, $\Psi_0 |_{\Phi=2\pi}$ and $\Psi_1 |_{\Phi=2\pi}$
will not be degenerate anymore, and the interaction $V$ gives rise to a
mixing term $\Psi_0^{\dagger} V \Psi_1$.
An explicit calculation shows that
$\Psi_0^{\dagger} V \Psi_1=
 \frac{4\Delta}{N} [ 1 - \cos2\pi/N ] \sim 8\Delta\pi^2/N^3$.
Therefore, we see that $\gamma= 3$ at $\Delta=0$ and $\Phi=2\pi$.
This agrees quite well with the numerical results of
Fig.\ (\ref{fig-gamma}) which were obtained for small systems of up to
$N=14$ sites and $M=7$ particles \cite{rem}.

%%%%%%%%%%%%%%%%%%%%%%%%%%%%%%%%%%%%%%%%%%%%%%%%%%%%%%%%%%%%%%%%%%%%%%%%
%
% Closing of the gap at the special fillings f=1/j
%
%%%%%%%%%%%%%%%%%%%%%%%%%%%%%%%%%%%%%%%%%%%%%%%%%%%%%%%%%%%%%%%%%%%%%%%%

\section{Closing of energy gaps at $1/\lowercase{r}$-filling}

Let us now study the behavior of the energy gap in the momentum $\pi$
sector away from half-filling.
As mentioned above, one generally expects the winding number of the
ground state to be $n=r$ for a $1/r$ filled band. As we have shown,
this is true for the half-filled band except at isolated values
of the interaction strength.
For fillings equal to $1/3, 1/4, 1/5, 1/6, \ldots$, we have
explicitly calculated the energy spectrum as a function of flux and
we find that the gap {\em closes} at $\Phi= \pi r$.
In Fig.\ (\ref{fig-3rdfill}), we show a plot of the $1/3$ filled case
with $\mu= 7\pi/12$ obtained by exact diagonalization of the H-I chain.
This interaction strength corresponds to a large
gap for the half-filled case. For $1/3$ filling, however,
the gap at $\Phi= 3\pi$ is closed.
Note that again various other degeneracies in the energy spectrum
occur at this flux value.
This behavior is unique to the $1/r$ filled chain --- and by particle-hole
symmetry to the $(r-1)/r$ filled chain. It does not
occur at all possible rational fillings such as, e.g., $2/5$.

The numerical results are again confirmed by perturbation theory.
Let us, e.g., look at the $1/3$ filled case with $M$ odd and $N= 3M$.
The ground state $\Psi_0$ is given as before and we need to identify
the state $\Psi_1$ which is degenerate with $\Psi_0$ at $\Phi=3\pi$.
We find
$\Psi_1|_{\Phi=0}=
 d_A^{\dagger} d_B^{\dagger} d_C^{\dagger} d_D d_E d_F \Psi_0|_{\Phi=0}$
with $A= -(M+5)/2$, $B= -(M+3)/2$, $C=-(M+1)/2$,
$D=(M-1)/2$, $E=(M-3)/2$ and $D=(M-5)/2$.

For $\Delta\neq 0$, the exact diagonalization implies that
$\Psi_0 |_{\Phi=3\pi}$ and $\Psi_1 |_{\Phi=3\pi}$ will remain degenerate.
Due to the structure of $\Psi_1$, we see that the mixing term
indeed vanishes in first order in $V$.
The second order mixing term is given by
\begin{eqnarray}
\lefteqn{\frac{1}{2} \Psi_0^\dagger V\cdot V \Psi_1 =} \nonumber \\
 & & \mbox{ }
 \frac{2 \Delta^2}{N^2}
 \sum_{\stackrel{{\scriptstyle i_1,i_2,i_3,i_4,}}{j_1,j_2,j_3,j_4}}
 \cos 2\pi(i_3-i_4)/N \; \Delta_{i_1+i_3,i_2+i_4} \cdot
 \cos 2\pi(j_3-j_4)/N \; \Delta_{j_1+j_3,j_2+j_4} \times \nonumber \\
 & & \mbox{ }
 \langle \Psi_0 |
 d^\dagger_{i_1} d^\dagger_{i_3} d^\dagger_{j_1} d^\dagger_{j_3}
 d_{i_4}         d_{i_2}         d_{j_4}         d_{j_2}
 d^\dagger_{A} d^\dagger_{B} d^\dagger_{C}
 d_D           d_E           d_F
 | \Psi_0 \rangle.
\label{eqn-scnd}
\end{eqnarray}
Explicitly performing the $576$ possible contractions, we find that
the diagrams arrange themselves into $36$ groups corresponding to
a given ordering of contractions over the indices $\{A,B,C, D,E,F\}$.
Only one index is left to be summed over and it is exactly this sum that
gives zero for all the $36$ groups individually.
Therefore, we have an exact cancellation of second
order diagrams for all $\mu$ in the critical region \cite{sp}, too.
A generalization of this calculation to other $1/r$ fillings is
straightforward, but becomes increasingly tedious as the number of
possible contractions increases as $(4+r)!$.

%%%%%%%%%%%%%%%%%%%%%%%%%%%%%%%%%%%%%%%%%%%%%%%%%%%%%%%%%%%%%%%%%%%%%%%%
%
% Conclusions
%
%%%%%%%%%%%%%%%%%%%%%%%%%%%%%%%%%%%%%%%%%%%%%%%%%%%%%%%%%%%%%%%%%%%%%%%%

\section{Conclusions}

We have presented numerical and perturbative evidence for degeneracies
in the momentum $\pi$ sector of the $1/r$ filled critical H-I model
on a finite ring.
Adding a non-integrable, say, next-nearest neighbor interaction
$W=\omega \sum_{j=1}^{N} n_j n_{j+2}$ to the Hamiltonian (\ref{eqn-ham}),
we find that {\em all} the degeneracies discussed above for
$\mu\neq\pi/2$ are lifted and therefore all gap-closings disappear.
The reader may now speculate that it is the integrability of the H-I chain
which leads to the closed gaps.
However, preliminary results from a related study of the repulsive Hubbard
chain show that there the gap closes at $U=0$ only.
It is an interesting question, if our results can be found also in other
one-dimnesional quantum systems such as the long-ranged Haldane-Shastry
chains.

Furthermore, we have shown by exact diagonalization and iteration of
the BA equations that the gap at $\Phi=2\pi$ ($P=\pi$) and half-filling
scales as power of the system size with variable exponent $\gamma$.
An analytical calculation of $\gamma$ is in preperation.

We close this paper by noting that as shown in \cite{rs94}, the here
considered structure of low-lying states in the H-I chain is
qualitatively the same in the SC model \cite{rs93}.
Therefore the results presented here will also hold in the SC model,
up to the renormalization of quantities such as the spin wave velocity
$v_s$.

\acknowledgments
The authors would like to thank B.\ Sriram Shastry, Joel Campbell and
Alexander Punnoose for many insights and fruitful discussions.
R.A.R.\ and H.-P.E.\ acknowledge financial support from the
Alexander von Humboldt foundation. H.-P.E. further acknowledges
support at Tours through the European Union's ``Human Capital and Mobility''
program.
H.-P.E.\ and B.S.\ benefited from a stay at the Aspen Center
for Physics.

% figures

\begin{figure}
  \caption{
  The full spectrum of the H-I chain for $N=12$ and $M=6$ at
  $\mu_3= 2\pi/3$. Note the level crossing for ground state and
  first excited state at $\Phi=2\pi$. Various other level crossings
  enhance the periodicity of the ground state such that its winding
  number is $n=6$.
  \label{fig-halffill}}
\end{figure}

\begin{figure}
  \caption{
  Energy of the ground state and first excited state and their difference
  at $\Phi=2\pi$ ($P=\pi$) for $N=8$ and $M=4$. Note the closing of the
  gaps at $\mu_2=\pi/2$, $\mu_3=2\pi/3$ and $\mu_4=3\pi/4$. The finite gaps
  in the regions $\mu_2\leq\mu\leq\mu_3$ and $\mu_3\leq\mu\leq\mu_4$
  are just visible.
  \label{fig-gap}}
\end{figure}

\begin{figure}
  \caption{
  Assuming $\Delta E \sim N^{-\gamma}$, we extrapolate $\gamma$ from
  finite size data of up to $N=14$ sites and half-filling.
  The error bars get larger as $\mu\rightarrow\mu_7$
  since we need $M\geq Q$ for $\mu\geq\mu_{Q-1}$.
  Data points for $\mu>\mu_7=6\pi/7$ are therefore not expected to
  obey the simple power-law behavior.
  \label{fig-gamma}}
\end{figure}

\begin{figure}
  \caption{
  The full spectrum for the H-I chain for $N=12$ and $M=4$ at
  $\mu= 7\pi/12$. Note the level crossing for ground state and
  first excited state at $\Phi=3\pi$ ($P=\pi$). Various other level
  crossings
  enhance the periodicity of the ground state such that its winding
  number is $n=6$.
  \label{fig-3rdfill}}
\end{figure}


\begin{references}
\bibitem{rem} We have also calculated the finite size behavior for
        the bandwidth at $\Phi=0$ and
	the numerical results reproduce the expected $1/N$ behavior
	with good accuracy even for these small system sizes.
\end{references}
\end{document}